%Paper: hep-ph/9312201
%From: JFGUCD@UCDHEP.UCDAVIS.EDU
%Date: Tue, 30 Nov 1993 23:42 PDT
%Date (revised): Thu, 2 Dec 1993 02:23 PDT

\input tables.tex
\input phyzzx.tex

\def\hpm{H^{\pm}}
\def\mhpm{m_{\hpm}}
\def\mhp{m_{\hp}}
\def\hp{H^+}
\def\hm{H^-}

\def\hsm{\phi^0}
\def\mhsm{m_{\hsm}}
\def\half{\ifmath{{\textstyle{1 \over 2}}}}
\def\h{h}

\def\tanb{\tan\beta}
\def\cotb{\cot\beta}

\def\wpm{W^{\pm}}
\def\wp{W^+}

\def\hpm{H^{\pm}}
\def\mhpm{m_{\hpm}}
\def\mt{m_t}
\def\mb{m_b}

\def\ptr{p_T}
\def\mtwo{M_{jj}}
\def\mthree{M_{bjj}}
\def\mfour{M_{bbjj}}

\def\delmw{\Delta\mw}
\def\delmt{\Delta\mt}

\def\ebtag{e_{b-tag}}
\def\emistag{e_{mis-tag}}

\def\ptr{p_T}

\def\ptmiss{\vec p_T^{\,\,miss}}

\def\gev{~{\rm GeV}}
\def\tev{~{\rm TeV}}

\def\fbi{~{\rm fb}^{-1}}

\def\madisonargonne{{\it Proceedings of the ``Workshop on Physics at Current
Accelerators and the Supercollider''}, eds. J. Hewett, A. White, and
D. Zeppenfeld, Argonne National Laboratory, 2-5 June (1993)}
%%Journal definitions

\def\prdj#1{{\it Phys. Rev.} {\bf D{#1}}}
\def\npbj#1{{\it Nucl. Phys.} {\bf B{#1}}}
\def\prlj#1{{\it Phys. Rev. Lett.} {\bf {#1}}}
\def\plbj#1{{\it Phys. Lett.} {\bf B{#1}}}

\def\ibid{{\it ibid.}}
\def\mt{m_t}
\def\wp{W^+}

\def\rta{\rightarrow}
\def\tanb{\tan\beta}

\def\cotb{\cot\beta}

\def\hsm{\phi^0}

\def\mhsm{m_{\hsm}}
\def\nsd{N_{SD}}

\def\mw{m_W}
\def\mz{m_Z}
\def\anti{\overline}

\def\ifmath#1{\relax\ifmmode #1\else $#1$\fi}
\def\half{\ifmath{{\textstyle{1 \over 2}}}}

\def\3quarter{{\textstyle{3 \over 4}}}

\def\ebtag{e_{b-tag}}
\def\emisid{e_{mis-id}}

\input phyzzx
\Pubnum={$\caps UCD-93-40$\cr}
\date{December, 1993}

\titlepage
\vskip 0.75in
\baselineskip 0pt
%\PHYSREV
\hsize=6.5in
\vsize=8.5in
\centerline{{\bf Detecting the $tb$ Decays of a Charged Higgs Boson at
a Hadron Supercollider}}
\vskip .075in
\centerline{ J.F. Gunion}
\vskip .075in
\centerline{\it Davis Institute for High Energy Physics,
Dept. of Physics, U.C. Davis, Davis, CA 95616}

\vskip .075in
\centerline{\bf Abstract}
\vskip .075in
\centerline{\Tenpoint\baselineskip=12pt%\parindent=1pc
\vbox{\hsize=12.4cm
\noindent We demonstrate that detection of a charged Higgs boson decaying via
$\hpm\rta t b$ will be possible at the LHC
in $gg\rta \hm t \anti b+\hp b\anti t$ events,
provided the $\hp\rta t\anti b$ coupling is substantial, $\mt\gsim 110\gev$,
and reasonably efficient and pure $b$-tagging can be performed.
}}

\vskip .15in
\noindent{\bf 1. Introduction}
\vskip .075in

Any extension of the simple one-doublet Higgs sector of the Minimal
Standard Model (MSM) that introduces either additional doublets or one or
more triplets (or both) necessarily implies that at least one pair of
charged Higgs bosons must exist.
Thus, a charged Higgs boson is the hallmark
of a truly non-minimal Higgs sector, and in particular of two-doublet models.
In contrast, the presence of more than one neutral Higgs boson could
be due to additional singlet Higgs representations beyond the single
MSM doublet. Consequently, the ability to detect a charged
Higgs boson is crucial to exploring a non-minimal Higgs sector.

Techniques for the detection of a charged Higgs boson at a hadron
supercollider are model-dependent.
Charged Higgs bosons that emerge from a model with two or more doublets
generally have substantial fermionic couplings.
All quarks of a given charge must couple to only one of the doublets
if we are to avoid flavor-changing neutral currents.
In the particularly simple and attractive
case of two-doublets, there are then only
two possible coupling patterns,
conventionally labelled as type-I and type-II.  In type-II models,
doublet 2 couples to up quarks while doublet 1 couples to down quarks.
\Ref\hhg{For a review see
J.F. Gunion, H.E. Haber, G.L. Kane, and S. Dawson, {\it The Higgs Hunter's
Guide}, Addison-Wesley, Redwood City, CA (1990).}\
Defining $\tanb\equiv v_2/v_1$, the ratio of the vacuum expectation values
of the neutral members of the two doublets,
the resulting $\hp\rta t\anti b$ coupling is given by
\foot{A $b$-quark mass of $\mb(2\mb)=4.7$ GeV is employed here and elsewhere
in our computations.}
$$ {g\over \sqrt 2\mw}[\mb P_R B+\mt P_L T]\eqn\coupspec$$
with $B=\tanb$ and $T=\cotb$,
where $P_{(R,L)}=\half(1\pm\gamma_5)$. The two-doublet
Higgs sector of the Minimal Supersymmetric Model (MSSM)
is necessarily of this type. In models of type-I, only doublet 2 has quark
couplings; doublet 1 couples only to vector bosons at tree-level.
The resulting $\hp\rta t\anti b$ coupling is obtained from Eq.~\coupspec\
by setting $T=-B=\cotb$.

Models with Higgs triplets alone cannot yield
$\rho\equiv\mw/(\mz\cos\theta_W)=1$
nor can they provide a source for fermion masses. The simplest model
that yields $\rho=1$ at tree-level consists of one $Y=1$ doublet,
one $Y=0$ real triplet and one $Y=2$ complex triplet.
\refmark\hhg\
The Higgs mass eigenstates include two charged Higgs bosons, one
of which ($H_3^+$ in the standard notation)
couples to $t\anti b$ according to Eq.~\coupspec\ with $T=-B=\tan\theta_H$,
where $\tan\theta_H$ characterizes the ratio of triplet vev's to the vev
of the neutral member of the doublet field.  The larger the role of
the triplets in giving mass to the $\wpm$ and $Z$, the larger $\tan\theta_H$.
The other charged Higgs boson of this model ($H_5^+$) couples strongly
to $\wp Z$ when $\tan\theta_H$ is large, and is generally
quite easily produced by $\wp Z$ fusion and detected in the $\wp Z$ final
state.\refmark\hhg\
This letter focuses on the seemingly
more challenging problem of developing techniques allowing the detection
of a charged Higgs boson with no tree-level $\wp Z$ coupling,
but with substantial fermionic couplings.

\REF\barnettetal{R.M. Barnett, R. Cruz, J.F. Gunion and B. Hubbard,
\prdj{47} (1993) 1048.}
\REF\dproyetal{D.P. Roy, \plbj{283} (1992) 403.}
In Refs.~\barnettetal\ and \dproyetal\ it was demonstrated that detection
of $t\rta\hp b$ decays, by searching for a violation of $\tau$-lepton
universality, would be relatively straightforward.
In contrast, it has generally been thought that detection of an $\hp$
which decays predominantly to $t\anti b$ would not be possible at a hadron
collider.  In this letter, we demonstrate that $b$-tagging
should allow detection of this decay mode for a substantial
range of top quark masses and $\hp\rta  t\anti b$ coupling strengths.
\Ref\preliminary{A preliminary report of the SSC results
using the techniques presented here
appears in J.F. Gunion and S. Geer, preprint UCD-93-32 (1993),
\def\madisonargonne{{\it Proceedings of the ``Workshop on Physics at Current
Accelerators and the Supercollider''}, eds. J. Hewett, A. White, and
D. Zeppenfeld, Argonne National Laboratory, 2-5 June (1993)}
to appear in \madisonargonne.}\

\vskip .15in
\noindent{\bf 2. Procedure and Results}
\vskip .075in

\REF\dgvi{J. Dai, J.F. Gunion and R. Vega, \prlj{71} (1993) 2699.}
The techniques employed are closely related to those developed for detection
of $t\anti t+\h$ with $\h\rta b\anti b$.\refmark\dgvi\
Indeed, the  $t\anti t b\anti b$
final state resulting from $gg\rta t\anti b \hm+\anti t b \hp$
(followed by $\hpm\rta t b$)
is precisely the same and the same backgrounds ---
$gg\rta t\anti t b\anti b$, $gg\rta t\anti t Z$ with $Z\rta b\anti b$,
and $t\anti t+jets$ --- must be brought under control. The large size
of the $t\anti t +jets$ background implies that the ability to tag three or
more
$b$-jets with good efficiency and purity will almost certainly be required.

Note that we shall generate signal and backgrounds as initiated by
$gg$ collisions.  If precisely three $b$-quarks are tagged, then
an alternative treatment can be envisioned
in which one studies $\hpm$ detection using the process
$g\anti b\rta \anti t \hp\rta \anti t t \anti b$ and its charge conjugate
for the signal and related $2\rta 3$ reactions for the backgrounds.
However, this $2\rta 3$ procedure obscures the fact that there really
is a second $b$-quark (associated with the colliding $\anti b$ above)
that could be tagged.  The $2\rta 4$ procedure adopted here
will better reflect the full kinematics of the underlying production
reaction in the types of high-transverse momentum configurations in
which it is probed; in particular, it will allow one of the $b$-tags
to be supplied by the $b$-quark which is an `invisible' spectator
in the $2\rta 3$ procedure.
It will also yield a more reliable representation
of the full complexity of the multi-jet environment in which one must
achieve the required isolations for multiple $b$-tagging. Finally,
it will more accurately include
combinatoric background effects.  The only short-fall of the $2\rta 4$
procedure is that it will generally underestimate the actual cross
section magnitudes.  This is well-understood
\Ref\magnitude{See for example, J.F. Gunion \etal, \npbj{294} (1987) 621;
F. Olness and W.-K. Tung, \npbj{308} (1988) 813;
D.A. Dicus and S. Willenbrock, \prdj{39} (1989) 751.}
as being due to the absence of the leading-log development of
QCD radiation from the final state $b$-quarks
in the tree-level $2\rta 4$ computation that is implicitly included
in defining the $b$-quark distribution employed in the $2\rta 3$ technique.
Thus, it will be necessary to multiply our $2\rta 4$ results by
significant QCD correction factors (to be specified below) in order
to reproduce correctly the cross section values.  These QCD
correction factors were estimated by simply comparing the uncut
$2\rta 4$ cross section with the uncut $2\rta 3$ cross section for the
reactions of interest.

Of course, the $2\rta 3$ cross sections will, themselves,
have higher order QCD corrections.
These are not available in the literature and are not computed here.
However, there is one component of the QCD corrections to the
$\hpm$ production process that will almost certainly emerge in
the full QCD-correction computation. Namely, it is reasonable to
anticipate that the QCD-corrected $\hpm$ production cross section
will approximately factorize into a fairly constant
overall $K$ factor (presumably
significantly larger than 1 for our $gg$-induced production mechanism,
but taken equal to 1 for the numerical results of this paper)
times the lowest order result with the $\hp\rta t\anti b$ coupling
expressed in terms of running quark masses. Indeed, in the closely related
computation of the QCD corrections to the $\hp\rta t \anti b$ width,
\Ref\qcdhtb{A. Mendez and A. Pomarol, \plbj{252} (1990) 461.
See, also, M. Drees and K. Hikasa, \plbj{240} (1990) 455, Erratum-\ibid\
{\bf B262} (1991) 497.}
replacing the physical quark masses by running masses yields an excellent
approximation to the full result. In the present computation
we refer all quark masses to their values at quark pair production
threshold, $m_f(2\mf)$.  This implies that
the running $t$-quark mass is slightly larger (smaller)
than the physical threshold $t$-quark
mass for $\mhp$ below (above) $2\mt$, whereas, since $\mhp$ is always much
larger than $2\mb$, the running $b$-quark mass is always significantly smaller
than the physical $b$ mass (typically by a factor of order 0.77).
Thus, in models of type-II,
inclusion of running mass corrections will significantly
decrease the $\hpm$ production cross section compared to its uncorrected
value when the $\mb$ term in the $\hp\rta t \anti b$
coupling is dominant, \ie\ when $\tanb$ is large.

We now give additional details on the precise procedures followed.
We use the MRS-D$0^\prime$ distribution functions
\Ref\mrs{A.D. Martin, W.J. Stirling, and R.G. Roberts, \plbj{306} (1993)
145; Erratum-\ibid\ {\bf B309} (1993) 492.} evaluated at a momentum
scale given by the subprocess center-of-mass energy.
We adopt an energy of $\sqrt s=16 \tev$
for the LHC. All jet and lepton momenta are smeared using
resolutions of $\Delta E/E=0.5/\sqrt E\oplus 0.03$ and
$\Delta E/E=0.2/\sqrt E\oplus 0.01$,
respectively ($\oplus$ means added in quadrature). An isolated lepton
($e$ or $\mu$) from one $t$ decay is used as the trigger.
The lepton is required to have
$\ptr>20\gev$, $|\eta|<2.5$ and to be separated by
$\Delta R >0.3$ from the nearest lepton or jet. A missing energy of
$|\ptmiss|>50\gev$ is required. At least three jets must be found in
the $|\eta|<2.5$ region with $\ptr>30\gev$.  (To be declared a jet,
a quark or gluon must be separated by $\Delta R>0.7$ from its nearest
neighbor.) Three tagged $b$-jets are then required. Only $b$-jets
(and, when mis-tagged, other jets)
with $|\eta|<2$ and $\ptr>20\gev$, isolated from
any other tagged jet by $\Delta R>0.5$, are considered.  Within these
kinematic restrictions, the probability for tagging a true $b$-jet
is taken to be that found in the SDC detector Technical Design Report,
\Ref\sdctdr{Solenoidal Detector Collaboration Technical Design Report,
E.L. Berger \etal, Report SDC-92-201, SSCL-SR-1215, 1992, p 4.15-4.16.}
which gives $\ebtag$ for the vertex detector
as a function of the $p_T$ of the $b$-jet. (Including tagging
of semi-leptonic $b$ decays via a lepton with
significant $\ptr$ relative to the main jet direction would
add to this efficiency.)
In this same kinematic range, the probability for mis-identifying a regular
gluon or light quark jet as a $b$-jet,
$\emisid$, is taken to be $0.01$, which is representative
of the values obtained in Ref.~\sdctdr, while the probability of
mis-tagging a $c$-jet is taken to be $0.05$.
We do not have available similar results for the LHC detectors.
It might prove that these $\ebtag$ and $\emisid$
values are optimistic given the multiple
interactions that occur in a given crossing when the LHC is run
at high instantaneous luminosity.

\FIG\hplus{$dN/d\mfour$ is plotted as a function of $\mfour$
for: the $gg\rta b\anti t \hp+\anti b t \hm$ signal (solid);
the $gg\rta t\anti t b\anti b$
background (dot-dash); and the $t\anti t g$ mis-tagged background (dashes).
For this plot we have employed the type-II two-doublet coupling
with $\tanb=1$, $\mt=140 \gev$, and an integrated
luminosity of $L=100\fbi$ at the LHC.
Signal curves are given for $\mhp=180$, $200$, $250$,
$300$, $400$ and $500\gev$.
Results do not include any QCD K-factors for the $t b \hpm$ signal
or $t\anti t b\anti b$ background.  No additional K-factor
for the $t\anti t g$ background is appropriate. QCD corrections
to the $\hp\rta t \anti b$ vertex are also not included.
For the solid signal curves, semi-leptonic decays of the $b$-quarks
are not included.  The effect of their inclusion is illustrated
in the case of the $\mhp=250\gev$ signal by the dotted histogram.}
\topinsert
%\vskip 2.65in
\vbox{\phantom{0}\vskip 5.0in
\phantom{0}
\vskip .5in
\hskip -0pt
\special{ insert user$1:[jfgucd.hplus]hplus_mfour.ps}
\vskip -1.45in }
\centerline{\vbox{\hsize=12.4cm
\Tenpoint
\baselineskip=12pt%\parindent=1pc
\noindent
Figure~\hplus: $dN/d\mfour$ is plotted as a function of $\mfour$
for: the $gg\rta b\anti t \hp+\anti b t \hm$ signal (solid);
the $gg\rta t\anti t b\anti b$
background (dot-dash); and the $t\anti t g$ mis-tagged background (dashes).
For this plot we have employed the type-II two-doublet coupling
with $\tanb=1$, $\mt=140 \gev$, and an integrated
luminosity of $L=100\fbi$ at the LHC.
Signal curves are given for $\mhp=180$, $200$, $250$,
$300$, $400$ and $500\gev$.
Results do not include any QCD K-factors for the $t b \hpm$ signal
or $t\anti t b\anti b$ background.  No additional K-factor
for the $t\anti t g$ background is appropriate. QCD corrections
to the $\hp\rta t \anti b$ vertex are also not included.
For the solid signal curves, semi-leptonic decays of the $b$-quarks
are not included.  The effect of their inclusion is illustrated
in the case of the $\mhp=250\gev$ signal by the dotted histogram.
}}
\endinsert

Additional cuts delineated below
tend to require that the second $W$ from $t$ decay must decay hadronically,
and that this $W$ combine with a tagged $b$-jet to form
a top quark. More precisely, the invariant mass of each pair of
jets, $\mtwo$, is computed and at least one pair {\it not containing any
tagged $b$-quark} is required to have
$\mw-\delmw/2 \leq\mtwo\leq\mw+\delmw/2$. In addition,
each pair of jets satisfying this criteria is combined
with each of the tagged $b$-jets to
compute the three-jet invariant mass, $\mthree$.
$\mt-\delmt/2\leq\mthree\leq\mt+\delmt/2$ is then required for at least
one $bjj$ combination.
If mass cuts of $\delmw=15\gev$ and
$\delmt=25\gev$ are used, only a small fraction of
signal events are eliminated for the earlier-quoted
jet and lepton energy resolutions,
whereas the reducible backgrounds are significantly decreased.
Finally, a plot of the $\mfour$ mass distribution is made, where both
$b$'s are required to be tagged and the two $j$'s must {\it not}
have been tagged. Typical results (before applying QCD corrections)
for the $\mfour$ distribution are shown in Fig.~\hplus.
For $\mhp$ not too near the $\hp\rta t \anti b$ decay threshold,
clear signal peaks are seen for $\mhp\lsim 400\gev$ where the
$\hp$ production rate is significant. As $\mhp$ approaches $\mt+\mb$,
the $b$ quark from the decay of the $\hp$ only infrequently has sufficient
$\ptr$ to be tagged, and the signal peak deteriorates.

The signal results shown by the solid curves
in Fig.~\hplus\ do not include the semi-leptonic
decays of the $b$-quark. If no attempt to identify those tagged $b$-quarks
that decay semi-leptonically is made, semi-leptonic decays reduce
the number of events in the central charged Higgs peaks.  On average,
the reduction is of order 25\%, part of which reduction
is due to the failure to reconstruct $\mt$ within $\delmt$ for events
with a semi-leptonic decay of the relevant $b$.
A sample is shown by the dotted
curve for the $\mhp=250\gev$ signal. It could be that those tagged
$b$-quarks which decay semi-leptonically can be identified by
observation of the lepton within the jet. (This is certainly possible
for the $\mu$'s, and may also be possible a significant fraction
of the time for $e$'s since they tend to have visible $\ptr$
relative to the main jet axis.) In this case, the visible momenta
can be rescaled to reflect the lost neutrino momentum, so as
to better reconstruct the average underlying $b$-quark momentum.  We
will not attempt this procedure here.

The only important backgrounds, after all cuts, turn out
to be the $t\anti t b\anti b$ continuum QCD background, and $t\anti t g$
where the $g$ is mis-tagged (with 1\% probability) as a $b$-jet.
The $t\anti t Z$, with $Z\rta b\anti b$, background is much smaller than
either. For a mis-identification probability of $0.05$ for a $c$-quark,
the $t\anti t c \anti c$ background is also not significant.  Intuitively,
this can be understood by noting that the $t\anti t b\anti b$ and $t\anti t
c\anti c$ cross sections are not very different once the tagged $b$ or $c$
is required to have significant transverse momentum, whereas the probability
for tagging a $b$ is much higher (a factor of 6) on average. Finally,
we note that including semi-leptonic decays
leads to background distributions
that are about 10\% smaller than those illustrated in Fig.~\hplus.

Before the significance of such signals can be computed, we must apply
appropriate QCD correction factors as explained earlier. We have
employed a K-factor of 2 for the $gg\rta t \anti b \hm+\anti t b \hp$
signal, and the $gg\rta t\anti t Z$ and $gg\rta t\anti t b\anti b$ backgrounds.
The $t\anti t g$ background has been computed employing a cutoff of
$\ptr>30\gev$ for the final state $g$.  As explained in Ref.~\dgvi,
this yields a $t\anti t g$ total cross section that is 60\% of the $t\anti t$
leading order cross section.  Thus, by computing the $t\anti t g$ process
with this cutoff, the effective $K$ factor for $t\anti t$ production
of $K=1.6$ is reproduced. Finally, the corrections to the $\hp\rta t\anti b$
vertex due to the running of the quark masses are applied.

 \TABLE\lhclum{}
 \midinsert
 \titlestyle{\twelvepoint
 Table \lhclum: Number of $100\fbi$ years, $Y$,  (signal event rate, $S$)
 at LHC required for a
 $5\sigma$ confidence level charged Higgs
 signal in a 40 GeV wide mass interval
 centered about $\mhp$,
 assuming 3 jets are tagged as $b$'s and $\emistag=1.0\%$.
 $BR(\hp\rta t\anti b)=1$ and model-II couplings with $\tanb=1$ are assumed.}
 \bigskip
 \hrule \vskip .04in \hrule
 \thicksize=0pt
 \begintable
  $\mt=$ | $\mhsm$ |  150 &  170 &  200 &  250 &  300 &  400 \nr
  110 | $Y(S)$ |   4.7( 179) &   1.8( 151) &
   2.6( 196) &   2.8( 159) &   2.7( 115) &   6.6( 105) \cr
 $\mt=$ |$\mhsm$ |  180 &  200 &  250 &  300 &  400 &  500   \nr
  140 | $Y(S)$ |   0.6(  93) &   0.5( 115) &
   0.5( 121) &   0.6( 102) &   1.3(  89) &   2.1(  72)  \cr
 $\mt=$ |$\mhsm$ |  200 &  220 &  250 &  300 &  400 &  500   \nr
  180  | $Y(S)$ |  25.5( 243) &   0.3(  52) &
   0.2(  60) &   0.3(  71) &   0.6(  69) &   1.1(  65)   \endtable
\hrule \vskip .04in \hrule
 \endinsert
 \vskip 1in

To estimate the statistical significance, $\nsd$, of a charged Higgs signal we
compute the signal event rate $S$ by focusing on either two or four
10 GeV bins centered about $\mhp$ (generally four bins is optimum,
but for $\mhpm$ close to the $tb$ decay threshold two bins gives the
best statistical significance).  The combinatoric background
from the signal reaction itself is estimated using the bins
immediately beyond the central bins, and is then
subtracted from the central bins.  $S$ is then computed by summing
the remainder event rate over the two or four central bins.
The background $B$ in this same mass interval is computed
by summing all the background rates, including the combinatoric
signal background subtracted in obtaining $S$, over the same central bins.
We then compute $\nsd=S/\sqrt B$. The results, after including
semi-leptonic $b$ decays, are presented in Table~\lhclum,
in terms of the number of LHC $100\fbi$ years ($Y$) required to detect
a given charged Higgs signal at the 5 sigma level.
Also given is the signal event rate
($S$) for this number of years.  The corresponding background rate
can be obtained by the relation $B=(S/5)^2$.  Table~\lhclum\
assumes model-II coupling
with $\tanb=1$, and gives results for four bins of combined width 40 GeV.
For the lowest $\mhp$ values considered for each $\mt$, better results
are actually obtained if a two-bin, \ie\ 20 GeV, interval is employed.
For a 20 GeV interval we find $Y(S)=2.1(81)$, $0.4(60)$ and $5.6(65)$
for $[\mt,\mhp]=[110,150]$, $[140,180]$, and $[180,200]$, respectively.
All are improvements over the four-bin results of Table~\lhclum.
Overall, we see that a charged Higgs boson with substantial
$\hp\rta t \anti b$ coupling and mass not too close to the $t\anti b$
decay threshold can be readily detected at the LHC
using $b$-tagging.  Since the largest background is that from $t\anti t g$,
with the $g$ mis-tagged as a $b$-quark, the clarity of the $\hpm$
signals would be significantly increased by
improving the tagging purity to $\emistag\lsim 0.005$,

\FIG\tanbsurvey{We plot $g_{eff}^2$
as a function of $\tanb$ for $\mt=110$, $140$, and $180\gev$
in the case of model-II coupling with $\mb=3.6\gev$ (typical
of the running $b$-quark mass evaluated at moderate $\mhp$).}
\topinsert
%\vskip 2.65in
\vbox{\phantom{0}\vskip 5.0in
\phantom{0}
\vskip .5in
\hskip +20pt
\special{ insert user$1:[jfgucd.hplus]tanbsurvey.ps}
\vskip -1.45in }
\centerline{\vbox{\hsize=12.4cm
\Tenpoint
\baselineskip=12pt%\parindent=1pc
\noindent
Figure~\tanbsurvey: We plot $g_{eff}^2$
as a function of $\tanb$ for $\mt=110$, $140$, and $180\gev$
in the case of model-II coupling with $\mb=3.6\gev$ (typical
of the running $b$-quark mass evaluated at moderate $\mhp$).
}}
\endinsert

The actual $\hp \rta t \anti b$ coupling may be either greater or smaller
than that obtained in model-II with $\tanb=1$.  In model-II,
the effective coupling strength decreases rapidly as $\tanb$ increases,
reaching an $\mt$ (and $\mb$) dependent minimum in the $\tanb\sim 5-7$
region, and then rises rapidly as $\tanb$ increases further, thereby
leading to a greatly enhanced $\mb$ coefficient.
This behavior is illustrated in Fig.~\tanbsurvey, where we plot
the effective coupling strength defined by $g_{eff}^2\equiv
(\mt^2 T^2+\mb^2 B^2)/(\mt^2+\mb^2)$, the denominator being
the coupling strength squared for $T=B=1$, \ie\ $\tanb=1$.
\foot{We have checked numerically that the $\hpm$ cross sections
are closely proportional to $g_{eff}^2$. This means
that the squares of the scalar and pseudoscalar couplings (there is
no interference) of Eq.~\coupspec\ enter the cross section
with very similar weights.}
In Fig.~\tanbsurvey\ we use $\mb=3.6\gev$ which is a typical
value for the running $b$-quark mass at moderate $\mhp$.
Although not plotted,
it is apparent that $g_{eff}^2$ rises rapidly above 1 for $\tanb<1$.

This behavior can potentially
affect our results in two ways.  First, it could, in principle alter
$BR(\hp\rta t \anti b)$ which we have
assumed to be near unity in Table~\lhclum.  However, even at the minimum,
the $\hp \rta t \anti b$ coupling strength squared is far larger
than that for any competing SM fermion-pair channel,
and the branching ratio remains
near unity.  Second, and very crucial, since the $\hp$ production rate is
proportional to $g_{eff}^2$, $\nsd$
must be multiplied by $g_{eff}^2$, and $Y(S)$ in Table~\lhclum\ must be divided
by $g_{eff}^4(g_{eff}^2)$. Obviously, a statistically significant
charged Higgs signal cannot be achieved for $\tanb$ near the minimum
point. Viable signals may only be obtained for $\tanb\lsim 2$ and large
$\tanb$.  More quantitatively,
if $\mt\sim 140(180)\gev$ a 5 sigma signal is obtained
in $2-3$ LHC $100\fbi$ years for $\mhp\lsim400\gev$ if $\tanb\lsim 1.5(1.7)$.
And, for all three $\mt$ values a 5 sigma signal is obtained
in $2-3$ LHC years for $\mhp\lsim
300\gev$ (but not too near threshold) if $\tanb\gsim 30$.

The values of $\tanb$ that are of greatest interest depend upon the
larger model context.
In a non-supersymmetric two-doublet model of type-II, small values
of $\tanb$ tend to be excluded, for the $\mhp$ values considered here,
by the experimental upper limit on the $b\rta s\gamma$ branching ratio.
\Ref\btosgam{J.L. Hewett, \prlj{70} (1993) 1045; V. Barger,
M.S. Berger, and R.J.N. Phillips, \ibid\ {\bf 70} (1993) 1368.}
In the MSSM (for which the Higgs sector is required to be a
two-doublet model of type-II) loops involving charginos cancel against
loops involving the charged Higgs, and there is no significant constraint
on $\tanb$ coming from the experimental limit on $BR(b\rta s\gamma)$.
\Ref\bertolinietc{S. Bertolino, F. Borzumati, and A. Masiero, \npbj{294}
(1987) 321; S. Bertolini, F. Borzumati, A. Masiero, and G. Ridolfi,
\npbj{353} (1991) 591;
R. Barbieri and G.F. Giudice, preprint CERN-TH 6830/93
(1993);
J. Lopez, D. Nanopoulos, and G. Park, preprint CTP-TAMU-16-93
(1993);
N. Oshimo, preprint IFM 12/92 (1992).}\ Both $\tanb\lsim 2$
and $\tanb\gsim 30$ are allowed regions of parameter space, and in these
regions discovery of the $\hpm$ will be possible in the mode explored here,
assuming decays of the $\hp$ to chargino+neutralino states do not
decrease $BR(\hp\rta t\anti b)$ to a value significantly below 1.
We note that small $\tanb\lsim 2$ and very large $\tanb\gsim 40$
are the preferred
regions in GUT scenarios for the MSSM in which $\lambda_b=\lambda_{\tau}$
is required at the GUT scale.
\Ref\bbo{See, for example, V. Barger, M.S. Berger, and P. Ohmann,
preprint MAD/PH/798 (1993) and references therein.}

In the case of model-I couplings, the $\mb$ term is not enhanced at
large $\tanb$, and so discovery of the $\hp$ in the $t\anti b$ decay mode will
be restricted to $\tanb\lsim 2$ as described above for model-II.
For the triplet model outlined earlier, $H_3^+$ discovery
in the $t\anti t b\anti b$ final state will require $\tan\theta_H\gsim 0.5$.
However, the limit on $BR(b\rta s\gamma)$ will tend
to rule out $\tanb$ and $\tan\theta_H$ values in the above ranges, unless
the models are supplemented with additional new physics that
yields loop corrections which cancel against the charged Higgs loop.

\vskip .15in
\noindent{\bf 3. Conclusion}
\vskip .075in

In conclusion, detection of a charged Higgs boson
in the $gg\rta t \anti b \hm+\anti t b \hp\rta t\anti t b\anti b$
production/decay mode will be possible for a significant range
of $\tanb$ values in a two-doublet model of type-II, which is
the most attractive Higgs sector extension and, in particular, is that required
in the Minimal Supersymmetric Model. Indeed, the $\tanb$ regions
for which discovery is viable correspond precisely to those preferred
when the MSSM is considered in a GUT context. In contrast,
parameter choices allowing $\hpm$ discovery via the $t\anti t b\anti b$ mode
in type-I two-doublet models and
in the above-described triplet Higgs model
(with custodial SU(2) symmetry at tree-level)
tend to be ruled out by experimental limits on $BR(b\rta s\gamma)$.

The ability to perform $b$-tagging with good efficiency
and purity is crucial to the techniques developed here. It will
be important for the LHC detectors to optimize their ability to tag
$b$-quarks in the multi-event per collision environment that will
prevail for the high instantaneous luminosity required to achieve
integrated luminosities of order $100\fbi$.

\smallskip\centerline{\bf Acknowledgements}
\smallskip
This work has been supported in part by Department of Energy
grant \#DE-FG03-91ER40674
and by Texas National Research Laboratory grant \#RGFY93-330.
I am grateful to M. Barnett, H. Haber, and F. Paige
for helpful conversations. Some of the Monte Carlo generators employed were
developed in collaboration with J. Dai, L. Orr and R. Vega.

\smallskip
\refout
%\figout
\end